\documentclass[twocolumn,tighten]{aastex63}
\DeclareUnicodeCharacter{0301}{*************************************}
\usepackage{amsmath}
\usepackage{booktabs,array,dcolumn}
\usepackage{tabularx}
\usepackage{scrextend} 
\usepackage{graphicx,subfigure}
\usepackage{booktabs,array,dcolumn,comment}
\usepackage{tabularx}
\def\msun{$\rm M_{\sun}$}

\def\mstar{$M_{\star}$}
\newcommand{\hi}{\mbox{\sc{Hi}$\,$}}

\newcommand{\mhi}{$M_{\rm HI}\,$}
\newcommand{\hal}[1]{\ensuremath{{\rm H}\alpha}}

\defcitealias{s19}{S19}
\newcommand{\wl}{$W_{50}\,$} 
\newcommand{\wxx}{$W_{20}\,$}
\newcommand{\wx}{$W_{10}\,$}

\begin{document}

\shorttitle{TBTF-Field}
\shortauthors{Sardone {\em et al.}}

\title{Closing the Gap between Observed Low-Mass Galaxy \hi\ Kinematics and CDM Predictions}

\correspondingauthor{Amy Sardone; NSF Astronomy and Astrophysics Postdoctoral Fellow}
\email{sardone.4@osu.edu}

\newcommand{\OSU}{\affil{Department of Astronomy, The Ohio State University, 140 West 18th Avenue, Columbus, OH 43210, USA}}
\newcommand{\CCAPP}{\affil{Center for Cosmology and Astroparticle Physics, The Ohio State University, 191 West Woodruff Avenue, Columbus, OH 43210, USA}}
\newcommand{\OSUphys}{\affil{Department of Physics, The Ohio State University, 191 West Woodruff Avenue, Columbus, OH, 43210, USA}}
\newcommand{\IAS}{\affil{School of Natural Sciences, Institute for Advanced Study, 1 Einstein Drive, Princeton, NJ 08540}}
\newcommand{\Rutgers}{\affil{Department of Physics \& Astronomy, Rutgers, the State University of New Jersey, 136 Frelinghuysen Rd, Piscataway, NJ 08854}}
\newcommand{\CCA}{\affil{Center for Computational Astrophysics, Flatiron Institute, 162 Fifth Avenue, New York, NY 10010}}
\newcommand{\UBC}{\affil{Department of Computer Science, Math, Physics, \& Statistics, University of British Columbia, Kelowna, BC V1V 1V7, Canada}}
\newcommand{\CSIRO}{\affil{CSIRO Space \& Astronomy, Parkes Observatory, P.O. Box 276, Parkes NSW 2870, Australia}}

\author[0000-0002-5783-145X]{Amy Sardone}
\OSU \CCAPP

\author[0000-0002-8040-6785]{Annika H. G. Peter}
\OSU \CCAPP \OSUphys \IAS

\author[0000-0002-0372-3736]{Alyson M. Brooks}
\Rutgers \CCA

\author[0000-0003-4810-7803]{Jane Kaczmarek}
\UBC \CSIRO

\begin{abstract}

Testing the standard cosmological model ($\Lambda$CDM) at small scales is challenging. Galaxies that inhabit low-mass dark matter halos provide an ideal test bed for dark matter models by linking observational properties of galaxies at small scales (low mass, low velocity) to low-mass dark matter halos. However, the observed kinematics of these galaxies do not align with the kinematics of the dark matter halos predicted to host them, obscuring our understanding of the low-mass end of the galaxy-halo connection. 
We use deep \hi{} observations of low-mass galaxies at high spectral resolution in combination with cosmological simulations of dwarf galaxies to better understand the connection between dwarf galaxy kinematics and low-mass halos.  
Specifically, we use \hi{} line widths to directly compare to the maximum velocities in a dark matter halo, and find that each deeper measurement approaches the expected one-to-one relationship between the observed kinematics and the predicted kinematics in $\Lambda$CDM. We also measure baryonic masses and place these on the Baryonic Tully-Fisher relation (BTFR). Again, our deepest measurements approach the theoretical predictions for the low-mass end of this relation, a significant improvement on similar measurements based on line widths measured at 50\% and 20\% of the peak. Our data also hints at the rollover in the BTFR predicted by hydrodynamical simulations of $\Lambda$CDM for low-mass galaxies.

\end{abstract}


\section{Introduction}
\label{sec:intro}

While the standard cosmological model used to predict the structure in the Universe ($\Lambda$CDM $=$ cosmological constant $+$ Cold Dark Matter) works well on large scales, testing its validity on small scales is challenging.  Galaxies are the key to testing dark matter on small scales because they inhabit dark matter halos as low in halo mass as $M_h \approx 10^8 M_\odot$, corresponding to halo velocity scales of $v \approx 10$ km/s \citep{jethwa2018,kim2018,nadler2020}.  Tests of dark matter depend on relating galaxies to the virial mass and dark matter distribution of the halos that surround them \citep[cf.][]{buckley2018}.   

For galaxies with $M_h \gtrsim 10^{12} M_\odot$ or $v \gtrsim 200$ km/s, the galaxy-halo connection can be measured directly by gravitational lensing. Another method to determine the galaxy-halo connection is statistical, to match the number density, clustering, and assembly history of an ensemble of galaxies to the simulated statistics of dark-matter halos \citep[``abundance matching'' or ``halo occupation distribution", e.g.,][]{berlind2002,tasitsiomi2004,behroozi2013,moster2021}, which has been used to match galaxies to halos down to $M_h = 10^{10} M_\odot$ \citep[see][for a review]{wechsler2018}. Another method, and the primary one to push below $M_h  = 10^{10} M_\odot$, is to match galaxies to halos based on the kinematics of the galaxies themselves.  Galaxy kinematics, and in particular the peak of the galaxy rotation curve $v_{max}$ \citep[or some proxy thereof;][]{lelli2019}, are used to observationally link galaxies at fixed $v_{max}$ to  dark-matter halos of fixed $V_{max}$, the peak of the halo circular velocity curve.  

However, for decades, it has been difficult to reconcile the kinematics observed in dwarf galaxies with the kinematics of the dark matter halos predicted to host them in the $\Lambda$CDM cosmological model \citep{kuzio2008,ferrero2012,Miller2014}.  This is apparent in the velocity function (VF), the number density of galaxies (or halos) as a function of velocity, with the observed VF of dwarf galaxies consistently much lower than the VF predicted for cold dark matter (CDM) halos, by about a factor of five at velocities of $\sim 30-40$ km/s, corresponding to total baryonic masses of about $10^8 M_\odot$ according to abundance matching \citep{zwaan2010,2011ApJ...739...38P,2015MNRAS.454.1798K}. Suppression in the halo mass function on account of novel dark matter physics (like the free streaming of a warm dark matter candidate) may reconcile the galaxy-to-halo mapping \citep{zavala2009theory,2011ApJ...742...16T,schneider2017}.  Alternatively, a number of authors show that the H\textsc{i} surveys on which the VF is based have significant observational incompleteness and/or bias \citep[e.g.,][]{maccio2016,brooks2017,chauhan2019}.  While these solutions can bring the observed VF closer to the completeness-adjusted theoretical VF, they do not necessarily address kinematic discrepancies in observations of individual dwarf galaxies.

Kinematic discrepancies in individual dwarf galaxies manifest themselves in one of the most fundamental relations in astronomy, the Tully--Fisher (1977) relation, between a galaxy's luminosity, which can be translated into stellar mass, and a measure of the galaxy's rotational velocity. This relation breaks down for gas-rich galaxies with rotation velocities below 100 $\rm km \, s^{-1}$, but is resolved once the gas mass is included in the relation, known as the Baryonic Tully-Fisher Relation \citep[BTFR;][]{mcgaugh2000}. 
Because of the steepness of the stellar-to-halo mass relation (SHMR) expected from abundance matching for halos with $V_{max} < 40$ km/s (baryonic masses of $10^8M_\odot$), the BTFR is expected to roll over, such that galaxies spanning a wide range of baryonic masses should inhabit halos of roughly identical mass \citep{sales2017,dutton2017,mcquinn2022}.  However, many of the measurements at low masses show that galaxies still follow the power-law BTFR \citep{begum2008tf,bradford2015, lelli2019} extrapolated from higher masses, or have even smaller velocities than expected based on the extrapolation \citep{papastergis2015}.
As such, rotation velocities observed in the smallest galaxies are systematically lower than expected for their baryonic masses, and in comparison to the velocities of the halos expected to host them.

The most promising avenue for reconciling the VFs and galaxy-halo mapping of individual galaxies is understanding the mapping between the halo velocity $V_{max}$ and the observable rotation velocity in galaxies.  For gas-rich galaxies, the rotation velocity of the gas disk can be measured either by integral field spectroscopy (of either the neutral H\textsc{i} component via the 21-cm line, the ionized component in the optical, or the molecular component in mm wavelengths) or by spatially unresolved measurement of the the 21-cm line flux.  Rotation velocities traced by \hi{} typically extend far beyond the stellar disk, and have a better chance of reaching the maximum halo velocity \citep{lelli2019}. VFs are typically constructed with the latter, using line widths measured at 50\% of the peak flux (\wl) of the 21 cm neutral hydrogen line (\hi{}) as a proxy for $2v_{max}$, with the HIPASS \citep{Meyer2004} and ALFALFA surveys \citep{2005AJ....130.2598G}.  

\wl works well for massive galaxies with the ``double horn" spectral feature, a result of the fact that much high-column material rotates at $v_{max}$.  However, most rotation curves are still rising at the last detected \hi{} point for dwarf galaxies with baryonic masses below $M\sim 10^8 M_\odot$, leading to a Gaussian \hi{} profile \citep[e.g.,][]{begumMNRASfiggssurvey,hunter2012,ott2012,mcquinn2022}.  
Analysis of high-resolution hydrodynamic simulations with CDM cosmology show that the \hi{} line width measured at \wl{} severely underestimates the maximum halo velocity $V_{max}$, a problem that becomes more extreme for lower-mass systems \citep{brook2016,maccio2016,yaryura2016,brooks2017,dutton2019}. 
Galaxies can be matched to halos by fitting the rotation curves with a mass model, but typically with strong assumptions about the shape of the density profile \citep{ferrero2012,brook2015tbtf,read2016,read2017,katz2017,TrujilloGomez2018,li2020halomodel,mcquinn2022}.  Cored halo models fit the best, and some fits point toward a reconciliation of the VF \citep{li2019late,mcquinn2022}.  However, inferring the density profile and halo mass of the smallest dwarf galaxies is challenging because of the thickness of the gas disk, non-circular motions, and non-equilibrium physics \citep{stilp2013global,stilp2013turbulence,oman2019,roper2022}, and the shape of the halo density profile is still not fully understood.

In this work, we explore the idea that spatially unresolved but highly spectrally resolved measurements of halo kinematics by single-dish telescopes may offer a better opportunity to measure luminous matter near the peak of the halo circular velocity function (the peak being relatively insensitive to the details of the halo density profile).  Single-dish observations are more sensitive to the low gas column densities we expect deep into the dark-matter halo of dwarfs.  Moreover, these dishes can probe diffuse gas at even quite large angular scales.  Single-dish observations have been shown to detect the extended, diffuse gas throughout the circumgalactic medium of larger galaxies \citep{das2020,sardone2021}.  Thus, deep single-dish observations can improve measurements of the kinematics from the \hi{} line taken at \wl{} for low-mass galaxies.  Measurements taken at 20\% of the peak line flux (\wxx), as compared to \wl{}, in low-mass galaxies from hydrodynamic simulations better probe the maximum velocity in a simulated dark matter halo \citep{maccio2016,brooks2017,dutton2019}.  Even deeper measurements may allow us to probe farther into the wings of the Gaussian line profiles of dwarfs, potentially allowing for an even better match between \hi{} and halo kinematics.

Here we show that accurate measurements of the HI line width in the lowest-mass galaxies (baryonic masses less than $10^8 M_\odot$) tightens the constraints on their rotation velocities. 
We find that 1) the shape of the HI line profile is significantly affected by the spectral resolution and the robustness of the detection, 2) precise measurements of the profile at 20\% and 10\% of the peak intensity better probe the dynamics of the dark matter halo than measurements at 50\%, 3) these measurements help to reconcile the discrepancy between low mass galaxy kinematics and the observed velocity function by including the lowest density \hi{} in the line width measurement, and 4) that precise measurements of \hi\ line widths can reduce the scatter in rotation velocities typical for extremely low mass galaxies along the BTFR, and approach the rollover in the BTFR expected in simulations of galaxies in CDM cosmologies. 

This paper is organized as follows. We describe our sample selection for our observations, data reduction, and simulations in Section \ref{sec:sample}. The methods we used for our analysis are laid out in Section \ref{sec:methods}. We present our main results in Section \ref{sec:results}, and discuss these results in the context of previous and future work in Section \ref{sec:discussion}. Our main takeaways can be found in Section \ref{sec:summary}.  

\bigskip
\section{Sample}
\label{sec:sample}

Our target sample consists of matched-property \hi{} observations and simulations. The details of both are given below. 

\subsection{Observational Selection}
Our sample of seven targets (see Table \ref{sampletab}) was chosen to be made up of low-mass galaxies with previous detections in both \hi\ and optical. Each of the seven galaxies has a low signal-to-noise ratio (SNR) detection within the footprint of the \hi\ Parkes All Sky Survey \citep[HIPASS;][]{barnes2001,koribalski2004}. The low SNR detection within HIPASS is a natural result of the all-sky design of the survey, which did not prioritize detections of very low mass galaxies. We required our galaxies to have previously measured \wl{} rotation velocities of $v_{rot} < 40 \, \rm km \, s^{-1}$. Each galaxy was also required to be nearby (less than 10 Mpc), and outside the virial volume of more massive systems. These requirements give us the best chance of observing galaxies that have not been affected by environmental processes and can be measured more accurately within the Parkes $14.1'$ beam. Previously measured \hi\ masses (\mhi) of these galaxies fall between \mhi $= 10^{6-7.6}$ \msun. 

\begin{deluxetable}{lcccr}
\tabletypesize{\footnotesize}
\tablecaption{Observed Galaxies}
\renewcommand{\arraystretch}{1.25}
\tablewidth{0pt}
\tablehead{
\colhead{Source} &
\colhead{RA} &
\colhead{Dec} &
\colhead{Dist} &
\colhead{$M_{B}$}  \\
\colhead{} &
\colhead{[deg]} &
\colhead{[deg]} &
\colhead{[Mpc]} &
\colhead{}
}

\decimalcolnumbers
\startdata
ESO349-031 & 2.056 & -34.578 & 3.18 & -11.9\\
DDO6 & 12.455 & -21.015 & 3.33 & -12.3\\
ESO199-007 & 44.517 & -49.383 & 6.03 & -12.5\\
AM0106-382 & 17.091 & -38.21 & 8.2 & -12.96\\
DDO226 & 10.766 & -22.247 & 4.97 & -13.63\\
ESO300-016 & 47.544 & -40.003 & 8.79 & -14.2\\
ESO553-046 & 81.774 & -20.678 & 6.7 & -14.7\\
\enddata

\tablecomments{Distances obtained from m-M magnitudes listed in the NASA/IPAC Extragalactic Database (NED) where all measurements are based on TRGB distances apart from ESO300-016 and AM0106-382, which are Tully-Fisher distances.}
\label{sampletab}
\end{deluxetable}

\subsection{Observations and Data Reduction}
\label{sec:obs}

\begin{figure*}
    \centering
    \includegraphics[trim=0cm 0cm 0cm 0cm, clip, width = 2.1\columnwidth]{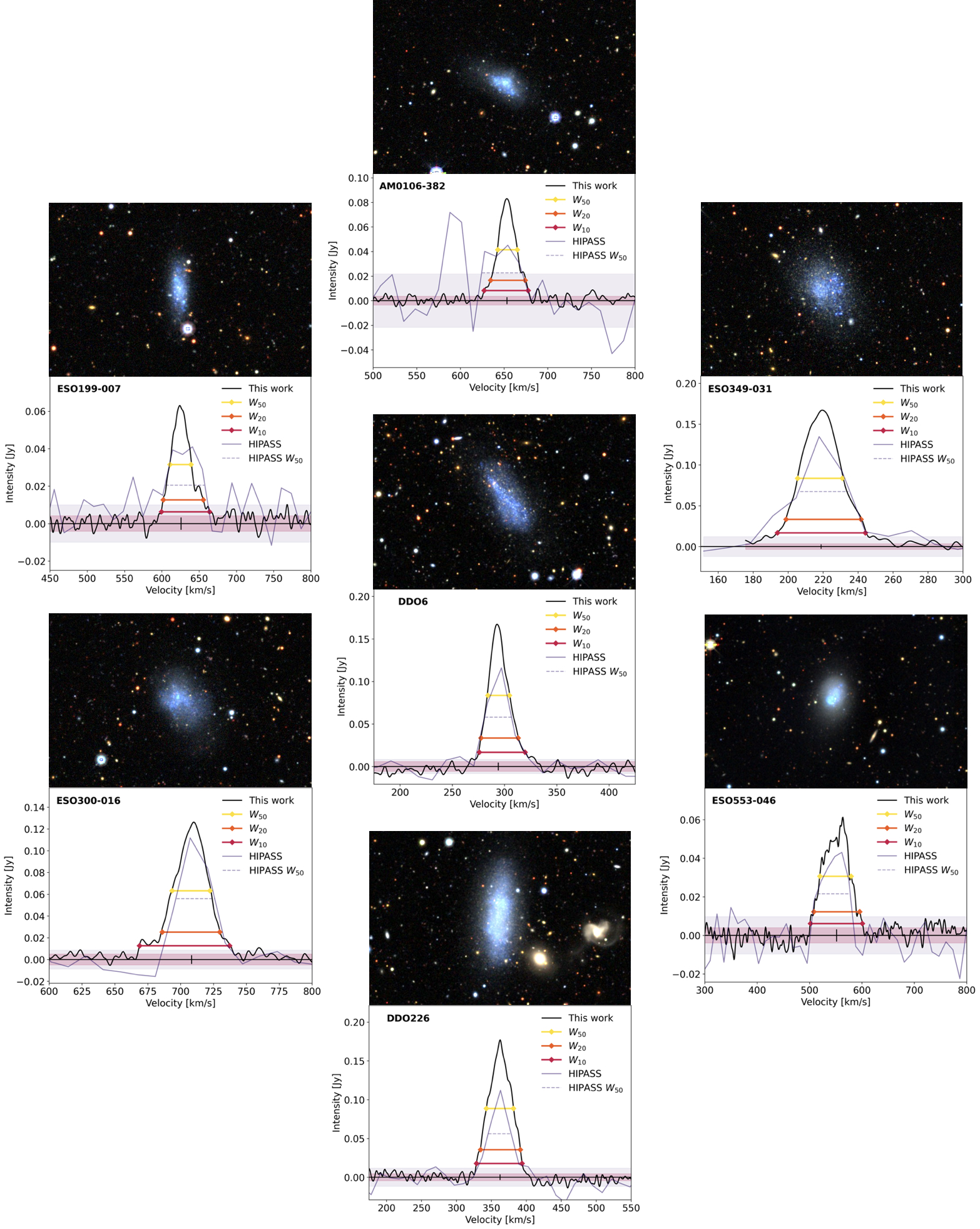}
    \caption{Total integrated HI spectra and optical images. Spectra from this work are shown in black where the line width at 50\%, 20\%, and 10\% of the peak intensity are shown in yellow, orange, and red, respectively. Spectra from the HIPASS survey are shown in light purple, where the dashed line represents the measured 50\% line width. The $1\sigma$ noise estimates are represented as shaded regions where HIPASS is in light purple and this work is in pink. The small vertical line represents the galaxy's central velocity based on a Gaussian fit to our data. Optical images were extracted from the DESI Legacy Survey DR9 \citep{dey2019} and have been stretched for optimal display.}
    \label{spectra}
\end{figure*} 

We were awarded 18 hours on the Parkes radio telescope for this program, P1047, to observe the 21-cm line of neutral hydrogen of seven low mass field dwarf galaxies (see Fig.~\ref{spectra}) using the Ultra Wide-bandwidth, Low-frequency receiver \citep[UWL;][]{2020PASA...37...12H}. We recorded dual-polarisation data over $2^{15}$ channels with a frequency resolution $0.82 \, \rm km \, s^{-1}$ (3.9 kHz) in sub-band 5, which covers a frequency range of 1344--1472 MHz. 
Each galaxy was observed for around 2.5 hours, using the On-Off calibration method. In practice, we pointed the telescope on-source for 2 minutes and moved to an off position three beam widths away along right ascension from the on-source position for an equal amount of time. We used the flux calibrators PKS~1934-638 and PKS~0407-658 at the start of each observing run and collected data for 180 seconds on these calibrators. 
The data were collected between September 3--13, 2020. We assumed a $T_{sys} = 22$ K, and a gain of $0.8 \, \rm K \, Jy^{-1}$. Using an average line width of $\sim 25 \, \rm km \,  s^{-1}$, we determined a theoretical sensitivity limit of 1 mJy, or 0.8 mK.
 This sensitivity corresponds to an \hi\ mass sensitivity of \mhi{} $\simeq 3 \times 10^5$ \msun.

We reduced the data using the SDHDF\footnote{\url{https://bitbucket.csiro.au/projects/cpda/repos/sdhdf_tools}} spectral line processing software. We followed this using our own Python scripts to perform continuum fitting, baseline subtraction, and spectral smoothing using a Gaussian kernel. In order to make an accurate measurement of the global \hi{} line profile, we required both excellent spectral resolution and sensitivity from our observations of each galaxy. We measured the galaxy spectra at a native spectral resolution of 0.82 $\rm km s^{-1}$ and with a measured $1 \sigma$ rms noise from 3.4 to 5.7 mJy (median of 4.2 mJy). We used a 2nd degree polynomial to fit and subtract the baseline in our spectra. The Parkes UWL receiver is stable on the frequency scales that are important to us, and therefore the spectra did not need higher order fitting. We smoothed the spectra using a Gaussian kernel of size two. 

We also obtained spectra from the HIPASS\footnote{\url{https://www.atnf.csiro.au/research/multibeam/release/}\label{note1}} public data platform, which have a final velocity resolution of 13.2 $\rm km s^{-1}$.  We note here that the HIPASS survey was an all sky survey, and was not designed to detect such low mass galaxies. From the spectra, we generated total integrated \hi{} line profiles which we used to compare to our integrated line profiles (Fig.~\ref{spectra}).

\subsection{Simulations}
\label{subsec:sims}

The simulated dwarf galaxies are drawn from the ``Marvel'' and ``DC Justice League'' (DCJL) suites of simulations, which yield a sample of over 200 simulated dwarf galaxies.  A full description of the simulations can be found in \citet{Munshi2021}, but we describe the relevant points here.  The Marvel simulations adopt a force resolution of 60 pc, and gas, initial star, and dark matter masses of 1410 \msun, 420 \msun, and 6650 \msun, respectively).  The Marvel galaxies are representative of Local Volume dwarf galaxies, being $\sim$2-7 Mpc away from a massive Milky Way-like neighbor.  The DCJL dwarfs are drawn from the ``Near Mint'' resolution runs, which have 170 pc force resolution, and gas, initial star, and dark matter masses of $2.7\times10^4$ \msun, 8000 \msun, and $4.2\times10^4$ \msun, respectively.  The DCJL simulations are centered on Milky Way-mass galaxies and their local environments.  The dwarfs here are isolated, found outside of the virial radius of the Milky Ways at $z=0$ but within $\sim$1 Mpc.

To match the observational sample, the simulated dwarfs were selected to be isolated (not within the virial radius of another galaxy) at $z=0$ and to have stellar masses $6.4 <$ log(M$_{\star}$/\msun) $< 7.4$.  These criteria yielded 15 dwarf galaxies from Marvel, and an additional 4 from DCJL \citep[from Sandra and Ruth,][]{Akins2021}, for a total of 19 simulated dwarf galaxies.  Star formation rates (SFR) measured over the last 100 Myr yield log(SFR) values between $-3.80$ and $-2.17$ for the sample, with $B$-band magnitude from $-10.5$ to $-14.5$.  The $V_{\rm max}$ values of the sample range from 28 to 50 km s$^{-1}$, and \mhi\ from $6\times10^6$ \msun\ to $3\times10^8$ \msun.

For each of the 19 dwarfs, we created mock \hi\ data cubes along three random sight lines.  The \hi\ mass fraction of every gas particle is calculated based on the particle’s temperature, density, incident cosmic UV background flux, self-shielding of H$_2$, and dust shielding of both \hi\ and H$_2$ \citep{Christensen2012}.  The mock \hi\ data cubes are designed to match the resolution of the observational sample, with a velocity resolution of 0.8 km s$^{-1}$.  Although not relevant for our unresolved \hi\ line widths, the data cubes were also created with the approximate spatial resolution of Parkes (corresponding to $\sim$20 kpc per pixel assuming the simulated dwarfs are at 5 Mpc).  The software to generate the cubes considers the line-of-sight velocity of each gas particle.  Galaxies are randomly oriented in the volume, so we chose our three random sight lines to correspond to the $x$, $y$, and $z$ axes of the simulation volume. In addition to the line-of-sight velocity, we add a thermal velocity component.  To account for the thermal velocity, the \hi\ mass of each gas particle is assumed to be distributed along the line-of-sight in a Gaussian distribution with a standard deviation given by the thermal velocity dispersion, $\sigma = \sqrt{{kT}{m_{\rm HI}}}$, where $T$ is the temperature of the gas particle and ${m_{\rm HI}}$ is the mass of the gas particle. 

Note that the data cubes are idealized, with no noise and no detection limits. Injecting artificial noise into the data cube at the same level as our observations would not affect the measurement made at $W_{10}$, and therefore the comparisons between the mock results and observations are not impacted by the lack of noise in our mock data cubes. From these mock \hi\ data cubes, $W_{10}$, $W_{20}$, and $W_{50}$ values are measured for the three random sight lines for each simulated dwarf galaxy.

\section{Methods and Analysis}
\label{sec:methods}

In this section, we provide the methods used to derive each galaxy parameter required for our analysis.

\subsection{Robust measurements of \hi\ line width}
To measure the line width of the HI emission from each target, both observed and simulated, we measure the value at 10, 20, and 50\% of the peak intensity of the emission, and calculate the width in km/s. These are shown as the red, orange, and yellow line widths in Fig.~\ref{spectra} and tabulated in Table \ref{derivedtable}. Measurements at 50\% of the peak are the most common in literature as that is often the best possible measurement in low SNR observations of dwarf galaxies in non-targeted \hi{} surveys. In order to ensure a robust measurement at 10 and 20\%, we required a SNR of at least 10 for our observed galaxies. Each of our targets met this criteria. The prevalence of low column density gas necessitates these deeper measurements, which is evident as we have measured increasingly larger line widths with each deeper measurement (Fig.~\ref{spectra}). This suggests that low mass galaxies, in particular galaxies with spectral profiles that tend toward Gaussian shapes rather than the double horned profiles of more massive galaxies, require deeper and higher spectral resolution measurements of their \hi{} line widths to obtain a more accurate understanding of the kinematics within their halos. 

Our measured $1\sigma$ noise estimate is represented by the pink shaded region in Fig.~\ref{spectra}. We calculated line width errors using an MCMC model fitting a Gaussian to the spectral line data. We found a fractional error from both the \wl{} and \wxx{} linewidths of around 2\%. 
We also compared our data with spectra from the HIPASS survey for each target (see Fig.~\ref{spectra}, purple). We performed the line width measurements in the same manner as specified above. We were only able to take measurements at 50\% of the peak in the HIPASS data (dashed line) as a result of both the low SNR ($1\sigma$ noise estimate represented by the purple shaded region), and low velocity resolution ($\sim 14 \, \rm km \, s^{-1}$ as compared to our $0.82 \, \rm km \, s^{-1}$). We followed the same method to estimate a fractional error of the \wl{} linewidth for the HIPASS data using the MCMC model, which resulted in a larger range of fractional errors with an average of 46\%. Each of these line widths are listed in Table \ref{derivedtable}.

\begin{deluxetable*}{lcccccccc}[t]
\tablecaption{Derived Properties of the Observed Galaxies}
\tabletypesize{\normalsize}
\renewcommand{\arraystretch}{1.25} 
\tablewidth{0pt}

\tablehead{
\colhead{Source} &
\colhead{$W_{10}$} &
\colhead{$W_{20}$} &
\colhead{$W_{50}$} &
\colhead{$W_{50}^{\rm HIPASS}$} &
\colhead{$V_{halo,max}$} &
\colhead{$log_{10}$M$_{\hi}$} &
\colhead{$log_{10}$M$_{*}$} &
\colhead{$log_{10}$M$_{halo}$} \\
\colhead{} &
\colhead{[km/s]} &
\colhead{[km/s]} &
\colhead{[km/s]} &
\colhead{[km/s]} &
\colhead{[km/s]} & 
\colhead{[\msun]} &
\colhead{[\msun]} &
\colhead{[\msun]} \\
\colhead{(1)} &
\colhead{(2)} &
\colhead{(3)} &
\colhead{(4)} &
\colhead{(5)} &
\colhead{(6)} &
\colhead{(7)} &
\colhead{(8)} &
\colhead{(9)} 
}
\startdata
ESO349-031 & 50.3 & 42.9 & 25.6 & 27.6 & 30 $\pm$ 2.1 & 6.94 & 6.41 & 9.86 \\
DDO6            & 43.7 & 35.5 & 20.6 & 25.4 & 32 $\pm$ 2.2 & 6.98 & 6.62 & 9.95 \\
ESO199-007 & 64.3 & 53.6 & 28.0 & 52.4 & 32 $\pm$ 2.2 & 7.02 & 6.63 & 9.96 \\
AM0106-382 & 50.3 & 39.6 & 22.3 & 46.2 & 34 $\pm$ 2.4 & 7.51 & 6.85 & 10.05 \\
DDO226        & 65.1 & 56.9 & 38.7 & 33.0 & 39 $\pm$ 2.7 & 7.48 & 7.27 & 10.23 \\
ESO300-016 & 68.4 & 43.7 & 28.9 & 28.8 & 38 $\pm$ 2.6 & 7.71 & 7.15 & 10.18 \\
ESO553-046 & 98.9 & 87.4 & 59.4 & 62.8 & 43 $\pm$ 3.0 & 7.23 & 7.6   & 10.37 \\
\enddata
\tablecomments{(1) Source name. (2) Line width measured at 10\% of the peak of the \hi{} line profile. (3) Line width measured at 20\% of the peak of the \hi{} line profile. (4) Line width measured at 50\% of the peak of the \hi{} line profile. (5) Line width measured at 50\% of the peak of the \hi{} line profile from the HIPASS data set. (6) The maximum velocity measured within the observed galaxy's dark matter halo, derived using Eq.~\ref{eq_vmax}. (7) \hi{} mass measured from the observed galaxies. (8) Stellar mass measured following prescription in Leroy+2019. (9) Mass of the halo using the SMHM relation from the observed stellar mass.}
\label{derivedtable}
\end{deluxetable*}

\subsection{Measuring $V_{rot}$}
\label{halomass}
Rotation velocities are derived from the HI line width and do not include an inclination correction due to the nature of our sample. Galaxies in our targeted mass range rarely have disk structure and often have puffy, irregular morphology, which makes an inclination estimate extremely uncertain \citep[cf.][]{el-badry2018}. As such, we have chosen not to introduce the large uncertainties associated with inclination corrections in these dwarfs and have instead adopted the following estimation of the rotation velocity for our observational data, which can be considered as a lower limit on the rotation velocity:

\begin{equation} \label{vrot}
V_{rot} = \frac{W_{50,20,10}}{2} \, .
\end{equation}

Rotation velocities for the mock \hi{} data were similarly measured from the \hi{} content in the simulated halos. To replicate the unknown inclinations of our observed sources, we measured the \hi{} line widths in the simulation from random inclinations within the cubes and derived $V_{rot}$ in the same manner. Each simulated profile was measured at 10, 20 and 50\% of the peak intensity. Since our simulations give us the opportunity to see this value from multiple orientations, we expect our observational measurements to reside within the spread of the simulated measurements.

\subsection{Probing the maximum halo velocity}
\label{methods-vmax}

We used different methods to estimate the maximum halo velocity for the simulations and observations, $V_{halo,max}$, in order to compare to the \hi-derived velocity. 
For our simulated galaxies, we extracted $V_{halo,max}$ directly from the simulation. 
For our observed galaxies, we used the stellar-to-halo mass relation (SHMR) from \citet{2010ApJ...710..903M,2013MNRAS.428.3121M} to derive halo masses for our observational data, which we then used to derive the maximum halo velocity.  We describe these methods further below. 

As described in Section \ref{subsec:sims}, we generated 19 mock \hi{} data cubes for simulated galaxies with stellar masses, magnitudes, and \hi{} masses within the same ranges as our observed galaxies.  For each simulated galaxy, the extracted $V_{halo,max}$ is based on the enclosed baryonic and dark matter content. That is, we calculate the circular velocity based on the mass enclosed, and find the maximum value. 

We use an alternate method to estimate $V_{halo,max}$ for our observed sample of galaxies by adopting the SHMR based on measured stellar masses for our galaxies. First we discuss how we estimate stellar masses, then we trace back to halo mass, and $V_{halo,max}$ for this method. 
We derive \mstar{} using unWISE cutouts \citep{lang2014} of reprocessed data from the WISE all-sky survey \citep{wright2010} following the methods prescribed in the z0MGS survey, including 0.15 dex uncertainties \citep[see the Appendix in ][for details]{leroy2019}. For our stellar mass calculation, we used tip-of-the-red-giant-branch (TRGB) distances when available \citep{lee1993}, and Tully-Fisher distances otherwise, and use the typically adopted distance uncertainty of 10\% for TRGB measurements, and 40\% for Tully-Fisher measurements (Table \ref{sampletab}). 
Although the SHMR predicts a probability distribution of stellar mass given a halo mass, we use Bayes' theorem to predict a probability distribution for each galaxy's halo mass given the stellar mass and its uncertainty \citep[as in][]{2022arXiv220909262G}.  We use the peak of the probability distribution for halo mass to estimate the maximum halo velocity.  
We define the halo mass specifically as the gravitationally bound mass within the virial radius ($M_{vir}$), defined as the radius enclosing an overdensity of 200 times the critical density of the Universe \citep{1996ApJ...462..563N}, and solve for the maximum velocity within the halo using eq. 10 from \citet{penarrubia2008} defined below:

\begin{equation} \label{eq_vmax}
    V_{halo,max} = V_{NFW(r_{max})} = \left[ \frac{G \, M_{vir}}{2 \, r_s} \frac {ln(3) - 2/3}{ln(1 + c) - c/(1 + c)} \right]^{1/2} ,
\end{equation}
\noindent where $r_{max} \simeq 2 r_s$, $r_s$ is the scale radius, and $c$ is the concentration parameter ($c \equiv r_{vir}/r_s$). A $1\sigma$ uncertainty in the $M_{vir}$ results in a $V_{halo,max}$ fractional error of about 7\%. The last pieces to solve for are $c$ and $r_s$, which we do using the following equations for the virial radius, and the concentration parameter from eq. 10 in \citet{klypin2011}:
\begin{equation}
    R_{vir} = \left(\frac{3 \, M_{vir}}{4 \, \pi \, 200 \, \rho_c}\right)^{1/3} \, ,
\end{equation}

\begin{equation}
    c(M_{vir}) = 9.60 \left (\frac{M_{vir}}{10^{12} \, h^{-1} M_{\odot}}\right)^{-0.075} .
\end{equation}

\noindent Here, $R_{vir}$ is the radius enclosing an overdensity by a factor of 200 times $\rho_c$, the critical density of the Universe, $G$ is the gravitational constant, and we use $h = 0.7$.

\begin{figure*}[t!]
    \centering
    \includegraphics[trim=0cm 0cm 0cm 0cm, clip, width = \textwidth]{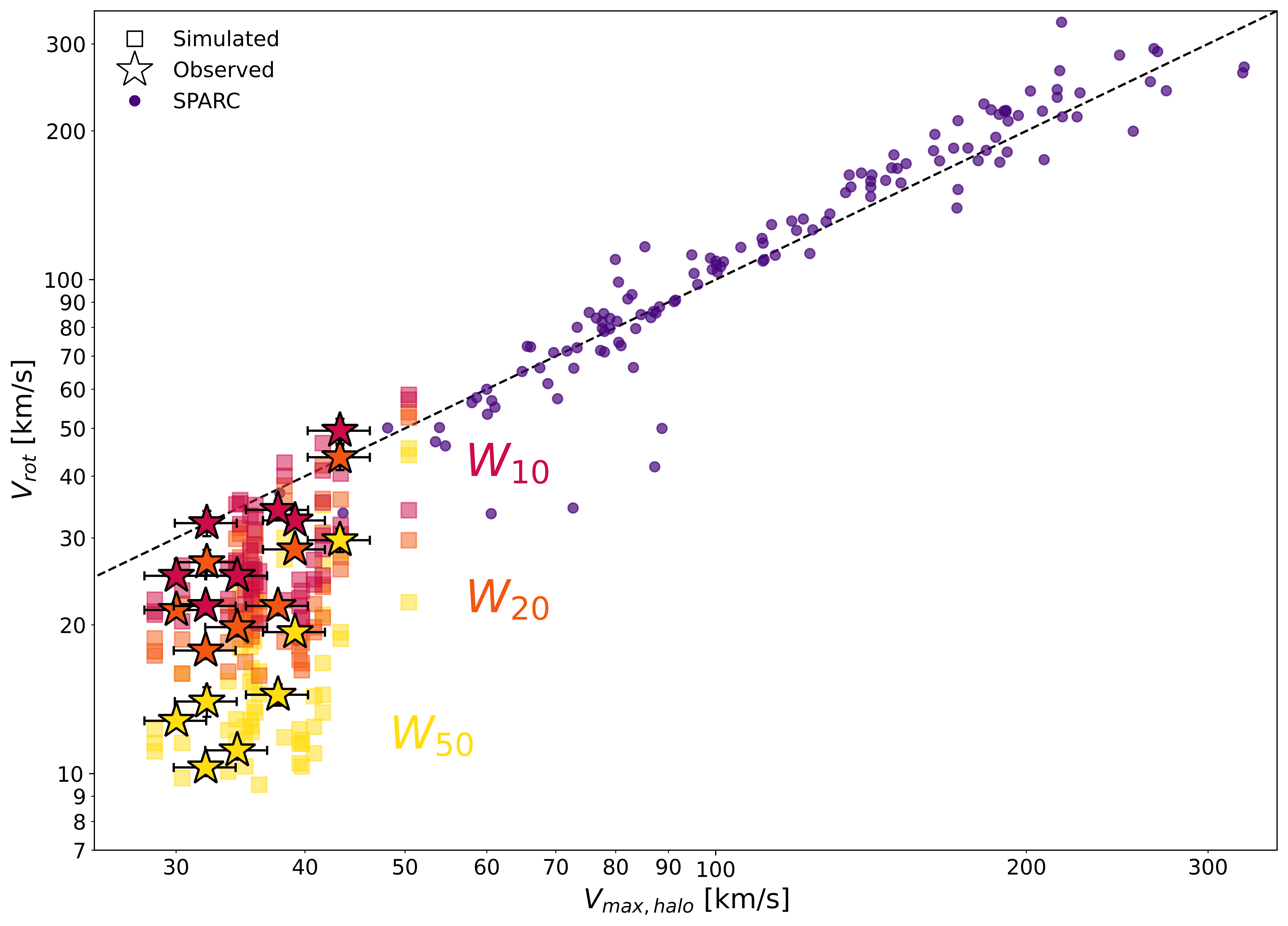}
    \caption{$V_{halo}-V_{rot}$ diagram. A direct comparison of the maximum velocity from dark matter in the halo to the maximum velocity measured from the \hi. Stars represent our \hi\ observations where $V_{halo}$ is derived as in Section \ref{methods-vmax}, and $V_{rot}$ is estimated from the \hi\ line widths measured at \wl, \wxx, and \wx. Squares represent our simulated data where $V_{halo}$ is the maximum velocity extracted directly from the dark matter, and $V_{rot}$ is estimated as in the observations but from mock \hi\ data cubes for three different sight lines for each dwarf. Purple circles represent data from the SPARC survey at a higher mass range. The dashed line represents a monotonic relationship between the rotation velocity measured from the \hi\ and the maximum velocity in the halo.}
    \label{vmax}
\end{figure*} 

\subsection{Mass measurements for the BTFR}
\label{btfrmethod}

The BTFR relates baryonic mass of a galaxy to the velocity of its rotation. We discussed rotation velocity in the sections above, as well as the stellar mass component of the baryonic mass ($M_{bary} = M_* + M_{gas}$). $M_{gas}$ in this case is the \hi{} mass corrected for helium by a factor of 1.33. We derive \hi{} masses by integrating under the global line profile, under the assumption that the \hi{} is optically thin, using:

\begin{equation}
    M_{\hi} = 2.36 \times 10^5 \, D^2 \int_{v_1}^{v_2} S(v) dv \, \rm M_{\sun} .
\end{equation}

\noindent We used distances, $D$ in Mpc, listed in Table \ref{sampletab}. The total integrated flux, $\int_{v_1}^{v_2}S(v) dv$ in Jy km $\rm s^{-1}$, encompasses the velocity range over which we see \hi{} emission. We estimate uncertainties on $M_{\hi}$ by taking the sum in quadrature with the distance and flux uncertainties. We use the method described in \citet{koribalski2004}, which is a modified version of that in \citet{fouque1990}, to derive our flux uncertainty.

\section{Results}
\label{sec:results}

To identify if spatially unresolved \hi{} line profiles can be used to further probe the kinematics of a dark matter halo and make comparisons to the expected relation on the BTFR, we first compare high spectral resolution and high SNR \hi{} line profiles from our data to the lower spectral resolution and lower SNR in the HIPASS data set. 
We then convert the \hi{} line width to a rotation velocity and compare this with the expected velocity of the associated dark matter halo. Finally, we use this same rotation velocity to place our galaxies along the BTFR and compare both to previous observations and theoretical expectations.

\begin{figure*}
    \centering
    \includegraphics[trim=0cm 0cm 0cm 0cm, clip, width = \textwidth]{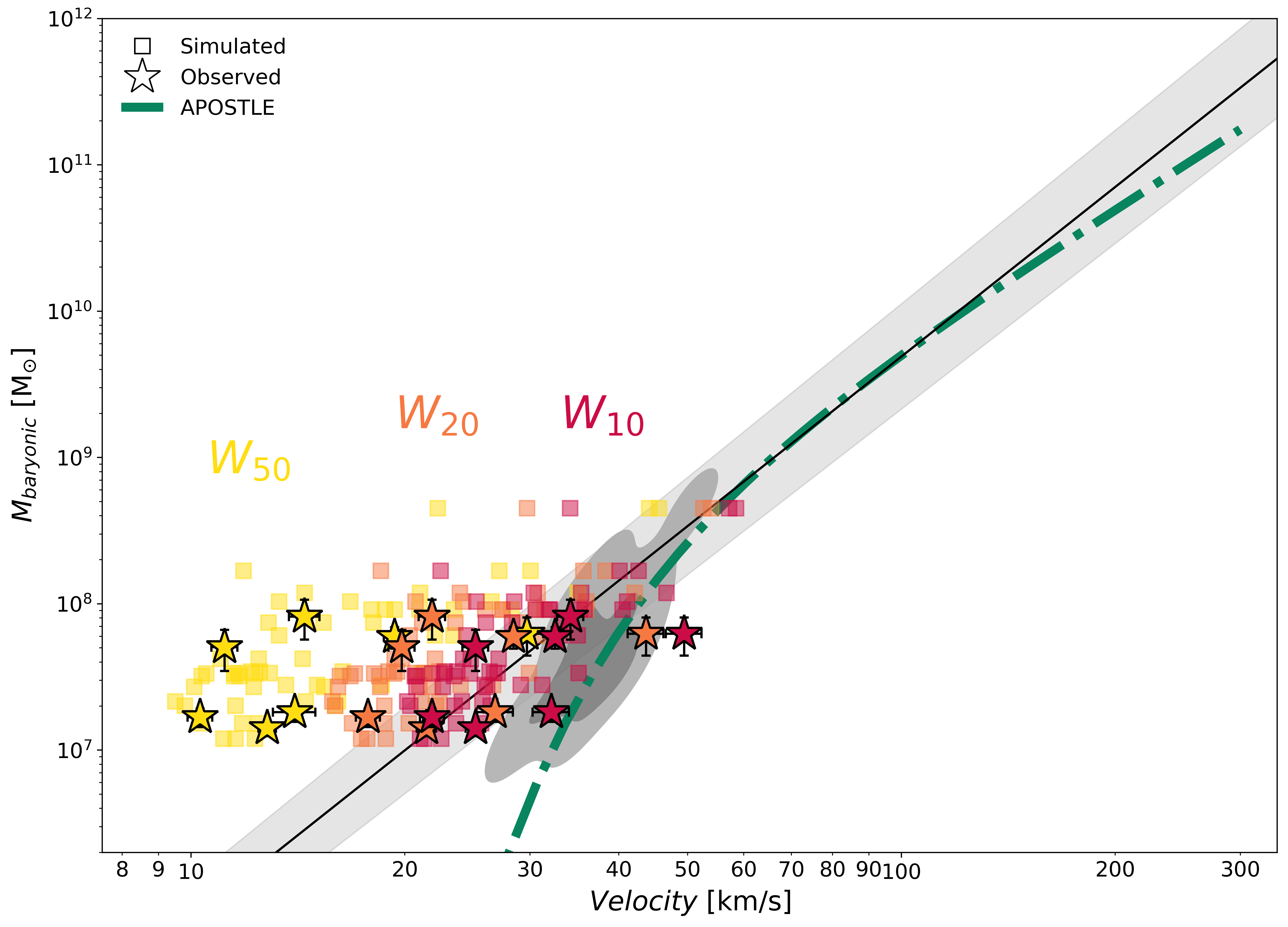}
    \caption{Baryonic Tully-Fisher Relation. Markers remain the same as Figure \ref{vmax} for observations and simulations. For the simulations, the gray contours represent the maximum velocity extracted directly from the enclosed mass, $V_{halo,max}$, as was plotted on the $x$-axis Figure \ref{vmax}. The solid black line with shaded error region is the linear fit and error region to the inclination-corrected SPARC data \citep{lelli2019}.    The green dash-dot line represents the predictions from the APOSTLE simulations \citep{sales2017}.  The ``true'' values from our simulations are in excellent agreement with the results in the APOSTLE simulations, both showing an expected turnover in the BTFR. Our deepest measurements (red) approach the simulated $V_{halo,max}$ (contours), but even our deepest \wx{} measurements (in both the simulations and the observations) do not demonstrate this same roll over.}
    \label{btfr}
\end{figure*} 

\subsection{Robust Measurements of the HI Line Width Probes Further Into the Maximum Halo Velocity}
\label{subsec:VF}

Theory predicts larger maximum halo velocities from low mass galaxies than those that have been previously measured by any galactic kinematic tracer \citep[e.g.,][]{kuzio2008,ferrero2012,Miller2014}. A comparison of our observed galaxies with synthetic observations of simulated galaxies shows why this is in fact expected in the context of $\Lambda$CDM. 

We plot \hi\ rotation velocities, $V_{rot}$, using \wl, \wxx, and \wx{} (see Section \ref{halomass}) from our observations with the $V_{halo,max}$ derived in Section \ref{methods-vmax}. These values are shown as stars in Fig.~\ref{vmax}.  The dashed line shows a one-to-one relation between $V_{rot}$ and $V_{halo,max}$, the goalpost we hope to attain if \hi{} line widths trace deeper into the dark matter halos to measure $V_{halo,max}$.  We also show rotation velocities from the slightly higher-mass SPARC survey \citep[purple points;][]{lelli2016}, whose rotation-curve-derived velocities were corrected for inclination effects, and present a robust example of galaxies along the one-to-one $V_{halo}-V_{rot}$ relation.  We find that $V_{rot}$ approaches the one-to-one relation between $V_{rot}$ and $V_{halo,max}$ (dashed line) as we measure deeper on the \hi{} line profile. This suggests that the deeper measurement of \wx{} is probing the low column density material seen in the wings of the line profiles, and this material is moving at a velocity approaching $V_{halo,max}$. 

Using our suite of fully cosmological hydrodynamic  simulations (see Section \ref{subsec:sims}), we plot the directly extracted maximum halo velocities, $V_{halo,max}$, and the \hi{} line widths from the associated mock \hi{} data cubes, again at \wl, \wxx, and \wx{} with Eq.~\ref{vrot}. These values are shown as shaded squares in Fig.~\ref{vmax}. Similar to the observed galaxies, we find that the rotation velocities derived from \wx{} are significantly closer to $V_{halo,max}$ than those from \wl. We note that the observations and simulations are in excellent agreement, both following the same trends when tracing deeper into the \hi{} line profile.  The agreement between simulations and observations suggests measurements low on the \hi\ profile may help to better reconcile the discrepancy observed in the low mass end of the velocity function of galaxies.

Using our simulated galaxies, we found on average that $V_{rot}$ derived from our simulated \wl{} probed 45\% of the maximum halo velocity based on the dark matter kinematic measurement, while \wxx{} probed 65\%, and \wx{} probed 77\%. These values can be inversely applied to say that the maximum halo velocity is a multiplicative factor of the rotation velocity estimated by the given the \hi{} line width as
\begin{equation}
    V_{halo} = 2.49 \substack{+1.35 \\ -1.38} \, \, V_{rot,W_{50}} \, ,
\end{equation}
\begin{equation}
    V_{halo} = 1.66 \substack{+0.79 \\ -0.72} \, \, V_{rot,W_{20}} \, , 
\end{equation}
\begin{equation}
    V_{halo} = 1.36 \substack{+0.56 \\ -0.5} \, \, V_{rot,W_{10}} \, . 
\end{equation}

\noindent These velocity corrections to the \hi{} line width may be applied to shallow archival \hi{} data to resolve discrepancies in the low mass end of the galaxy velocity function. 

We remind the reader that we consider the \hi{} rotation velocity to be a lower limit in both our observed and simulated data, since we are not correcting for inclination.  Despite neglecting inclination, the deeper \hi{} measurements close the gap in the low mass end of the $V_{halo}-V_{rot}$ relationship.  In fact, our deeper observations allow us to see the extrema in the gas velocity, which includes gas moving closer to $V_{max}$ along the line of sight, meaning that we are less reliant on inclination corrections to bring $V_{rot}$ into agreement with $V_{halo,max}$.

\subsection{Extremely Low Mass Galaxies Lie Below the BTFR}
\label{subsec:BTFR}

We use the same rotation velocities discussed in the previous section to populate the low-mass end of the BTFR in Fig.~\ref{btfr}. As before, we compare our low-mass galaxies with the well-characterized, higher mass, inclination-corrected galaxies from the SPARC survey \citep{lelli2016} using a fitted solid line to extend the relation down to lower masses. We use contours in Fig. \ref{btfr} to indicate the maximum velocity within the halo as measured directly from enclosed mass in the simulations. This is meant to serve as a goalpost if our derived rotation velocities are  tracing the maximum velocities within the galaxy halo as a function of baryonic mass. Both our observed and simulated \wl{}~data points lie to the left of the BTFR, as measured from inclination-corrected rotation curve fitting of higher mass galaxies, indicating that this is an insufficient measurement of the kinematics for these low mass galaxies. This result could be anticipated from Figure \ref{vmax}, where \wl{}~comes much closer to the one-to-one line with $V_{halo,max}$, but still lie systematically below it.

While the measurement at \wxx{} is much closer to the inclination-corrected SPARC BTFR fit \citep{lelli2019}, the \wx{} measurement lies along the BTFR fit and even on occasion crosses into the $V_{halo,max}$ contours. Like the observations, we note that our simulated \wx{} velocities derived from mock \hi\ data cubes agree well with the BTFR fit from the more massive galaxies.  However, their true $V_{halo,max}$ values show the same turn-down at low masses that have been predicted by other simulations within a $\Lambda$CDM context (e.g., NIHAO; \citet{dutton2017}, and APOSTLE; \citet{sales2017}).  This discrepancy will be explored further in future work, but hints at the possibility that observations may not be able to trace $V_{halo,max}$ in dwarf galaxies, even with deep data that probes further into the halo.

\section{Discussion}
\label{sec:discussion}

Since \hi{} is the most extended tracer of halo dynamics, commonly extending to much larger radii than optical tracers, we use deep observations of the \hi{} line widths in an attempt to measure the maximum rotation velocities of the dwarf galaxy dark matter halos. The data we obtained combines both high velocity resolution and high signal to noise to derive a deeper and more accurate measurement of the rotation velocity in very low mass galaxies. The combination of these two components (velocity resolution and SNR) is essential to understanding fundamental galaxy relations. Without a high SNR, we would not see the diffuse gas picked up by the single dish telescope that presents in the wings, deep into the line profile of the galaxy. Without high velocity resolution, we would smooth over the peaks and details in the profile that are used to measure mass and line width, resulting in inaccurate mass measurements and narrow line widths leading to low rotation velocities.  Lacking either of the two components leads to biases in the fundamental galaxy relations, and thus to misinterpretation of fundamental relations like the BTFR and the galaxy velocity function. 

Historically, low rotation velocities measured in low mass galaxies can be immediately attributed to two common types of measurements: 1) single dish measurements that lack either a high SNR or a fine spectral resolution, and 2) measurements from an interferometer at the farthest measurable radius in a galaxy, which can only detect structure smaller than some predefined angular scale as a natural consequence of the missing short spacings due to the telescope setup. This consequence of using an interferometer means it can completely miss any large scale structure moving with the galaxy, exacerbating the fact that the outermost detected \hi\ is nearly always on the rising part of the galaxy rotation curve in dwarfs and therefore not probing the maximum velocities in the galaxy. 

In this section, we discuss additional implications of our work in relation to previous observations and theoretical work.

\subsection{Change in Velocity with \hi\ Line Depth}
We see a clear progression to higher measured rotation velocities when moving from \wl{} to \wxx{} to \wx{}.  The fact that \wxx{} is larger than \wl{} has been noticed previously in both observations \citep[e.g.,][]{koribalski2004, bradford2015} and theoretical work \citep[e.g.,][]{brook2016, brooks2017, dutton2019}.  The increase in measured velocity is a natural consequence of the fact that the \hi\ line profile has a Gaussian shape in dwarf galaxies (see Figure \ref{spectra}).  In theory, as we measure further down on the \hi\ line profile we are probing the higher velocity gas that is further out in the rotation curve.  Our hope was that \wx{} would reach the maximum of the rotation curve.  We see in Figure \ref{vmax} that \wx{} comes closer to the expected $V_{halo,max}$ (for both observations and simulations), but the majority of points fall short of the one-to-one line (we discuss the impact of inclination corrections below). 

The excellent agreement between simulated and observed low-mass galaxies at high spectral resolution and sensitivity further bolsters the claim that dwarf galaxy kinematics are fully consistent with the CDM paradigm as modeled in the Marvel and DC Justice League simulations, and that the low \wl{} and \wxx{} measurements are a natural consequence of galaxy evolution on the dwarf scale.  The excellent match between the velocities and baryonic content of the simulated dwarfs lends them confidence for use in interpreting observations. 
Matching the BTFR means getting both the linewidths and the baryon content of simulated dwarfs right.  \citet{el-badry2018} showed that the FIRE simulations systematically underpredict both \wl{} and the gas fraction of dwarf galaxies, even as \wxx{} matches observed galaxies well.  In our simulations, not only do all three linewidth populations match well, but the cold baryon content of our galaxies does, too.

Because our observed and simulated data points are so well-matched, we argue that it is unlikely that $V_{halo,max}$ could be obtained with deeper data. The simulations are not limited in sensitivity, nor do we include mock noise.  Thus, they already offer as deep a probe of the \hi{} possible, yet the resulting \hi{} kinematics fail to trace $V_{halo,max}$.  This result suggests that \hi{} in dwarf galaxies simply does not reach the radius that traces the maximum rotation velocity. 

However, we note that any inclination correction could potentially move the data points to the right, increasing the velocities.  We discuss this further next.

\subsection{Inclinations}

Due to their shallow gravitational potential wells, very low-mass galaxies are extremely sensitive to baryonic feedback processes that can substantially redistribute their gas \citep[e.g.,][]{read2016, el-badry2017}.  Given the bursty nature of star formation in dwarfs \citep[e.g.,][]{mcquinn2010a, mcquinn2010b, weisz2012, kauffmann2014}, it is not clear that \hi{} morphologies and kinematics are traced by the stars, and thus it not clear that using the optically-derived inclinations is applicable to correcting the HI velocities.  Moreover, dwarf galaxies appear to become thicker relative to their size with decreasing mass \citep[e.g.,][]{Dalcanton2004,xu2023}, and stellar shapes may be a better tracer of the underlying (prolate) dark matter halo shapes in dwarfs \citep{Xu2020, Orkney2023}.  Both effects makes an inclination measurement determined from optical axis ratios more uncertain. \citet{mcquinn2022} note that the inclinations derived for dwarf galaxies in {\sc little things} \citep{hunter2012} via optical vs \hi{} morphologies have a mean offset of 8$^\circ$. 
Given these complications, we have presented our \hi{} velocities above without inclination corrections.  Likewise, \citet{el-badry2018} opted not to perform inclination corrections when comparing the FIRE simulated dwarfs to observed dwarfs, choosing instead to measure the simulations from random inclination angles for a more direct comparison, as we have also done.

\begin{figure}
    \centering
    \includegraphics[trim=0cm 0cm 0cm 0cm, clip, width = \columnwidth]{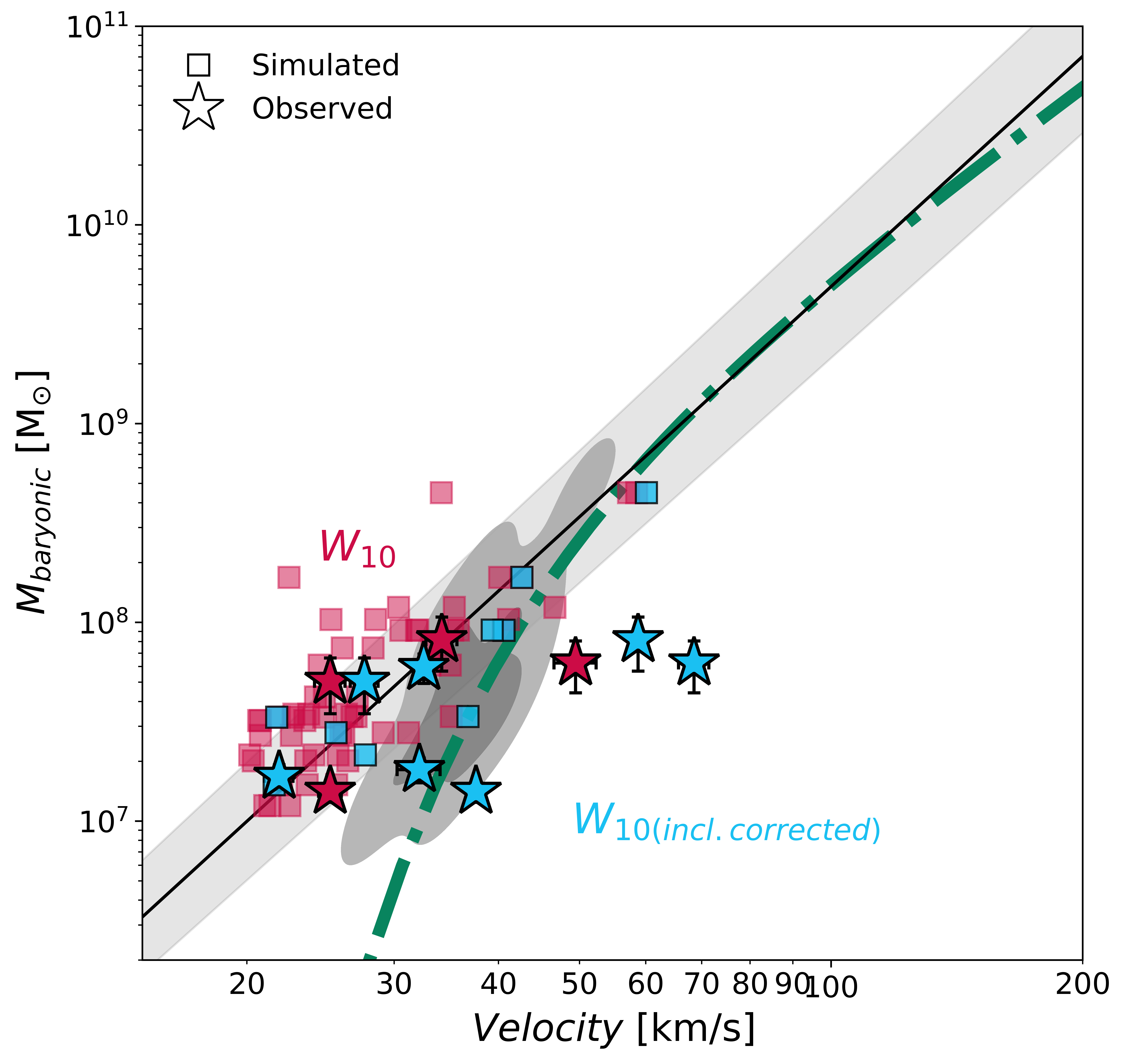}
    \caption{Baryonic Tully-Fisher Relation. Gray contours represent the maximum velocity in the simulated halos, $V_{max}$, as shown in Figure \ref{btfr}. The solid line and shaded error region is the linear fit and error region to the inclination-corrected SPARC data. Red stars are the rotation velocity as measured from the \wx{} line width, and blue stars have been corrected for inclination.  For the simulated dwarf galaxies, the non-inclination-corrected velocities are shown measured along three different sight lines (red squares).  A corresponding inclination-corrected velocity (cyan squares) is shown for the simulated dwarfs that could be oriented edge-on.}
    \label{btfr-incl}
\end{figure} 

Despite all of the above caveats, we show inclination corrected results in in Fig.~\ref{btfr-incl}. As we can see in Fig.~\ref{spectra}, the galaxies in our observed sample are generally puffy and show little clear structure. Estimating an inclination for this sample could introduce a large amount of uncertainty. Nevertheless, we correct our rotation velocities for inclination effects for the \wx{} observed data by dividing the line widths by $sin \,i$, using inclinations estimated from the optical light \citep[e.g.,][]{paturel1997} as listed in the HyperLEDA catalog \citep{2014A&A...570A..13M} and the Catalog of Local Volume Galaxies \citep{2019AstBu..74..111K}. In Fig.~\ref{btfr-incl} we compare the uncorrected velocities (magenta stars) with the inclination-corrected velocities (cyan stars). It can be seen that the distribution of velocities increases, highlighting the uncertainty introduced by inclination corrections. 
However, the inclination-corrected rotation velocities are generally higher than the non-inclination-corrected velocities, and we should therefore consider the non-inclination-corrected velocities as lower limits on the rotation velocity of the galaxy.  

As a further test of inclination corrections, we attempt to inclination correct the simulated dwarf galaxies.  We find that in nearly all of the dwarfs, the angular momentum vector of the central gas cannot be used to align the disk into an edge-on configuration, suggesting that the dwarfs are dispersion-dominated in most cases rather than rotation-dominated.  However, in some of the dwarfs we could identify a clear minor axis to the stellar component, and were able to rotate the galaxy so that it appeared edge-on.  We then repeated our generation of a mock \hi{} data cube and measured \wx{} for our edge-on galaxies.  We were able to do this for roughly half of the simulated dwarfs, with the remainder being too puffy to identify a minor axis.  The results are shown in Fig.~\ref{btfr-incl} as the cyan squares.  

Notably, \wx{} from the edge-on simulated dwarfs still underestimates $V_{halo,max}$ in many cases, in particular for the lowest baryonic masses.  In fact, in many cases the edge-on  \wx{} can be comparable to the non-inclination-corrected \wx{} values.  This confirms our earlier hypothesis that the \hi{} in the halos fails to trace the radius where maximum rotation occurs, at least for the simulated dwarfs with the lowest baryonic masses.

\subsection{Turnover in the BTFR}

Within CDM, we expect the BTFR to turn down at low velocities if the linewidths faithfully trace $V_{halo,max}$ because of the steep slope of the stellar-halo mass relation for $V_{halo,max} \lesssim 100$~km/s. In other words, galaxy formation is expected to become rapidly inefficient below halo masses of $\sim$10$^{10}$ M$_{\odot}$.   Despite the hope that \wx{} would trace $V_{halo,max}$, we do not see the expected turn-down in the inclination-uncorrected data.  It is notable, though, that we do not see the turn-down at \wx{} for either the observations or the mock \hi\ observations of the simulations, and yet we can determine from the simulations that $V_{halo,max}$ is still at higher velocities.  If the mock \hi\ observations had been able to trace $V_{halo,max}$ in the simulated dwarfs, we would have seen the turn-down in the BTFR. Recall that we do not impose a sensitivity or detection limit to our mock \hi\ data, nor do we include mock noise or other beam effects.  This suggests that the detectable \hi\ gas simply does not extend to the radius that traces $V_{halo,max}$.

\citet{mcquinn2022} measured rotation velocities at small radii for galaxies in a similar mass range, probing the kinematics in the inner part of the gravitational potential well. To help account for the small radii and the fact that the outermost measured velocities were still on the rising part of the rotation curves, they fit the rotation velocities assuming an underlying cored Einasto dark matter profile \citep{einasto1965} as motivated by simulated dwarf galaxies \citep{Lazar2020}.  This assumption allowed them to infer a maximum circular velocity for the best fit halo.  When placed on the BTFR, these models predict a turn-down that begins around 45 km/s at $M_{bary}$ $\sim 10^8$ \msun{} and steepens to $\sim 30-35$ km/s at $M_{bary}$ $\sim 10^7$ \msun, consistent with cold dark matter models and results from simulations that include baryon physics (e.g., the NIHAO \citep{dutton2017} and APOSTLE \citep{sales2017} simulations).  This steepening implies that galaxies with baryonic masses below $\sim 10^8$ \msun~ should populate dark matter halos with similar masses and circular velocities. Likewise, we see this same turn-down in Fig.~\ref{btfr} as the gray contour indicating the $V_{halo,max}$ values from our simulated dwarf galaxies.  Again, the fact that our deeper \wx{} measurements do not trace $V_{halo,max}$ suggests that even these deeper \hi\ measurements do not extend to a radius that traces the maximum circular velocity of the underlying dark matter halos. 

\subsection{Looking to the Future}

The correspondence between the measured linewidths from our deep, high-spectral-resolution data and the simulated dwarf sample suggests that we can accurately recover halo mass functions from shallower or lower-resolution survey data, even for very low mass dwarf galaxies.  Using these simulations, one can measure the correlation between linewidth and $V_{halo,max}$ for any specified HI survey, and then translate the linewidth-based velocity function in such a survey to a $V_{halo,max}$ function for dark-matter halos.  We have made a first measurement of such a relation in this work (Equations 6, 7, and 8). These relations enable the translation of an observationally-measured velocity function into theory space, which is especially important for the next generation of wide-field HI surveys and telescopes, e.g., WALLABY \citep{koribalski2020} or the SKA \citep{2009IEEEP..97.1482D}.

In parallel, our work can be used as a jumping off point to understand the connections among HI linewidths, galactic baryons, and their dark-matter halos more closely. Understanding the physical origin of both the scatter and the mean relation between the various metrics of linewidth and the halo velocity is important for more accurately inferring the former from the latter. Furthermore, it would be useful to obtain deep, high spectral resolution HI observations of low-mass galaxies that already have spatially resolved rotation curves, to determine if the halo velocities inferred by going deep into the lines matches those inferred from the spatially resolved but less deep data \citep[e.g., the samples from][]{li2020halomodel,mcquinn2022}. The new, deeper data could serve to extend the rotation curve as measured by the spatially resolved data, and inform us about the kinematic connection between the galaxy and its halo. 

Going beyond that, we would like to understand what physical processes shape the linewidths, and how linewidths and line shapes depend on the physical state of the galaxy (star-formation history, star-formation rate, gas fraction, etc.), and the physics of dark matter.  In Fig.~\ref{spectra}, our sample of seven galaxies show a diversity of line shapes, including one for which the line is far from symmetric (ESO553-046).  This galaxy is also the one with by far the highest star-formation rate, and thus the HI in this system may be far from dynamical equilibrium.  Furthermore, \citet{el-badry2018} suggests that the degree of dispersion support relative to rotational support can change the degree to which the line profiles look Gaussian.  Intriguingly, their toy model suggests that galaxies in cored dark matter halos, as would be expected in the case of strong baryonic feedback \citep[e.g.,][]{Governato2012, DiCintio2014, read2016} or if dark matter is self-interacting \citep{Vogelsberger2014, Robles2017}, should have a much more centrally peaked line profile than galaxies in cusped dark matter halos.  Thus, the diversity of linewidths has the power to tell us more than just the galaxy-halo connection, but can inform our understanding of the nature of dark matter and galaxy-evolution physics. 

Realizing this insight requires an expanded observed sample of low-mass galaxies, and a theory program that yields population statistics of galaxies for CDM and self-interacting dark matter models. For the future, we advocate for deep, high spectral resolution observations of the smallest gas-containing dwarf galaxies, and a matched suite of high-resolution simulations, which will be critical to providing a physics interpretation of the population properties of low-mass dwarfs.

\section{Summary}
\label{sec:summary}

The standard cosmological model used to predict the structure in the Universe is challenging to test on small scales. Theories of galaxy formation depend on tight observational constraints of the fundamental relationships among galaxies, namely, galaxy luminosity, mass, and kinematics. Thus far, we have been unable to resolve the tension between low mass galaxies and theories of galaxy formation, e.g., $\Lambda$CDM. This work reduces that tension by using \hi\ linewidths from sensitive, high spectral resolution data to match the baryonic kinematics of low mass galaxies to their dark matter halo kinematics. 

\begin{enumerate}
    \item We used deep, high velocity resolution \hi{} observations to get the clearest picture yet of the \hi\ line profiles of extremely low mass galaxies.  We derive rotation velocities by measuring the line width of the total integrated \hi{} spectral profile at 50\%, 20\%, and 10\% of the peak intensity. We have made robust (SNR $> 10$) measurements at \wl and \wxx, and \wx. This combination allows us to measure low column density \hi\ that is often hidden below the noise in observations of low mass galaxies.     We have measured increasingly larger line widths with each deeper measurement. This suggests that low mass galaxies, in particular galaxies with spectral profiles that tend toward Gaussian shapes rather than the double horned profiles of more massive galaxies, require deeper and higher spectral resolution measurements of their \hi{} line widths to obtain a more accurate understanding of the kinematics within their halos.
    \item We used mock \hi\ observations at a similar resolution of simulated dwarf galaxies to interpret the observed results.     We found a remarkable match between the simulated galaxy properties and the observed \hi{} properties, both in terms of mock observation linewidth results and baryonic content. The excellent match suggests that the simulations can be reliably used to interpret the observational results.  In the results below, we use the simulations in the to connect the observed \hi{} to the underlying dark matter halo properties of the observed galaxies.
    
    \item We use our deepest measurements of the \hi{} line width (\wx) to translate into a rotation velocity, and compare this to the expected maximum rotation from the dark matter halo. We find that the observed rotation velocities lie along the expected one-to-one line with an offset toward lower observed velocities. This implies that, while the \hi{} traces the dark matter kinematics better than any other observable, it is not a perfect one-to-one match. The fact that this result holds in our simulated dwarf galaxies suggests that even the deep, high-resolution  \hi{} observations are not tracing the maximum rotational velocities of the underlying dark matter halos. 
        \item Our rotation velocities measured from linewidths taken at \wx{} align with the Baryonic Tully-Fisher relation for low-mass galaxies. This is a significant improvement on previous work from both shallower measurements and rotation curve estimations.  However, we do not find the expected turn-down.  This is true in both our observed and simulated galaxies, yet we can determine from the simulated galaxies that a turn-down would be seen if we were able to find an observational probe of the true maximum circular velocity.  \end{enumerate}

The SKA pre-cursor survey WALLABY is already taking pilot observations.  Ultimately, the WALLABY survey will provide a new determination of the VF of dwarf galaxies \citep{koribalski2020}.  Our study outlines the inherent biases that exist in using the observed VF to test our theory of galaxy formation.  First,  shallower, lower SNR data prevent a measurement of the fullest linewidth, biasing the measured velocities to lower values.  Second, even the deepest \hi\ observations have not yet resulted in measurements of the expected maximum rotational velocities in dwarf galaxies.  This suggests that a comparison between observations and theory must be done in an apples-to-applies comparison in order to not introduce a second bias.  Future work with a larger observed dwarf galaxy sample and realistic simulations will elucidate the relation between observed dwarf galaxy kinematics and their underlying dark matter halos.

\acknowledgments
A.S. is supported by an NSF Astronomy and Astrophysics Postdoctoral Fellowship under award AST-1903834. A.H.G.P. is supported by NSF Grant No. AST-2008110. This research made use of the NASA/IPAC Extragalactic Database (NED). Resources supporting this work were provided by the NASA High-End Computing (HEC) Program through the NASA Advanced Supercomputing (NAS) Division at Ames Research Center. A.S. would like to thank Adam Leroy for his thoughtful feedback and conversations, and the OSU Galaxies group for helpful discussions. We would also like to thank George Hobbs for assistance with the new instrument reduction software. 

\bibliography{mybib.bib}
\bibliographystyle{aasjournal}

\end{document}